\documentclass[twocolumn,preprintnumbers,aps,superscriptaddress,nofootinbib]{revtex4-2}
\pdfoutput=1
\usepackage{amsmath,amssymb,nccmath,graphicx,color,hyperref}
\definecolor{darkblue}{rgb}{0.0,0.0,0.7}
\definecolor{nicered}{rgb}{0.7,0.1,0.1}
\definecolor{nicegreen}{rgb}{0.0,0.4,0.0}
\hypersetup{colorlinks,citecolor=nicegreen,linkcolor=nicered,urlcolor=nicered}
\graphicspath{ {figs/} }
\makeatletter
\newcommand{\oset}[3][0ex]{%
  \mathrel{\mathop{#3}\limits^{
    \vbox to#1{\kern-2\ex@
    \hbox{$\scriptstyle#2$}\vss}}}}
\makeatother
\newcommand{\ep}{\epsilon}
\newcommand{\CS}{\textcolor{darkblue}}
\newcommand{\mint}[1]{\scalebox{0.85}{#1}}
\newcommand{\Graph}[2]{\vcenter{\hbox{\includegraphics[scale=#1]{#2}}}}
\newcommand{\FDiag}[2]{
\begin{minipage}{0.12\textwidth}
$\includegraphics[width=\textwidth]{#2}\hspace*{-14ex}\raisebox{-1.5ex}{\CS{#1}}\hfill$
\end{minipage}
\hspace*{2ex}
} 

\allowdisplaybreaks

\begin{document}

\preprint{MSUHEP-21-011, P3H-21-035, TTP21-013}

\title{Fermionic corrections to quark and gluon form factors in four-loop QCD}

\author{Roman N. Lee}
\affiliation{Budker Institute of Nuclear Physics, 630090 Novosibirsk, Russia}

\author{Andreas von Manteuffel}
\affiliation{Department of Physics and Astronomy, Michigan State University,
East Lansing, Michigan 48824, USA}

\author{Robert M. Schabinger}
\affiliation{Department of Physics and Astronomy, Michigan State University,
East Lansing, Michigan 48824, USA}

\author{Alexander V. Smirnov}
\affiliation{Research Computing Center, Moscow State University,
119991, Moscow, Russia}
  
\author{Vladimir A. Smirnov}
\affiliation{Skobeltsyn Institute of Nuclear Physics of Moscow State University,
119991, Moscow, Russia}

\author{Matthias Steinhauser}
\affiliation{Institut f{\"u}r Theoretische Teilchenphysik,
Karlsruhe Institute of Technology (KIT),
76128 Karlsruhe, Germany}

\begin{abstract}
  \noindent
  We analytically compute all four-loop QCD corrections
  to the photon-quark and Higgs-gluon form factors involving
  a closed massless fermion loop.
  Our calculation of non-planar vertex integrals
  confirms a previous conjecture for the
  analytical form of the non-fermionic contributions to the
  collinear anomalous dimensions of quarks and gluons.
\end{abstract}

\maketitle


\section{Introduction}

Two of the most important processes which are studied in great detail at the
CERN LHC are the production of lepton pairs and Higgs bosons.
The total cross sections for Drell-Yan lepton pair production
through virtual photons and Higgs boson production through the dominant
gluon fusion channel are known to next-to-next-to-next-to-leading order (N$^3$LO) in
perturbation theory~\cite{Anastasiou:2015vya,Mistlberger:2018etf,Duhr:2020seh}
in the limit of an infinitely heavy top quark.
Historically, first the virtual
corrections have been computed and the real radiation contributions have been
added later.  In this paper we take an important step towards the N$^4$LO
corrections and provide analytic results for the fermionic contribution to the
virtual corrections, both to the Drell-Yan and Higgs boson production processes.

\begin{figure*}
\begin{center}
\FDiag{$n_f C_F^3$}{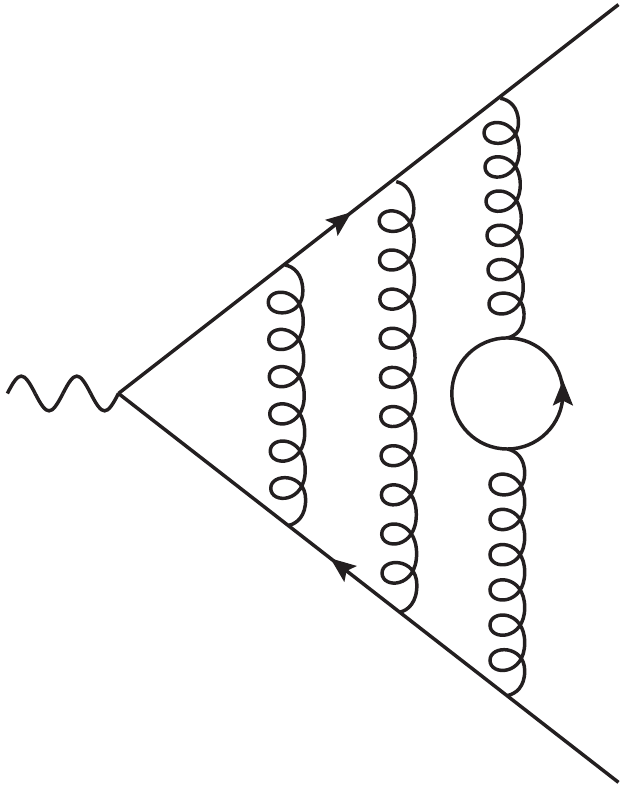}
\FDiag{$n_f C_F^2 C_A$}{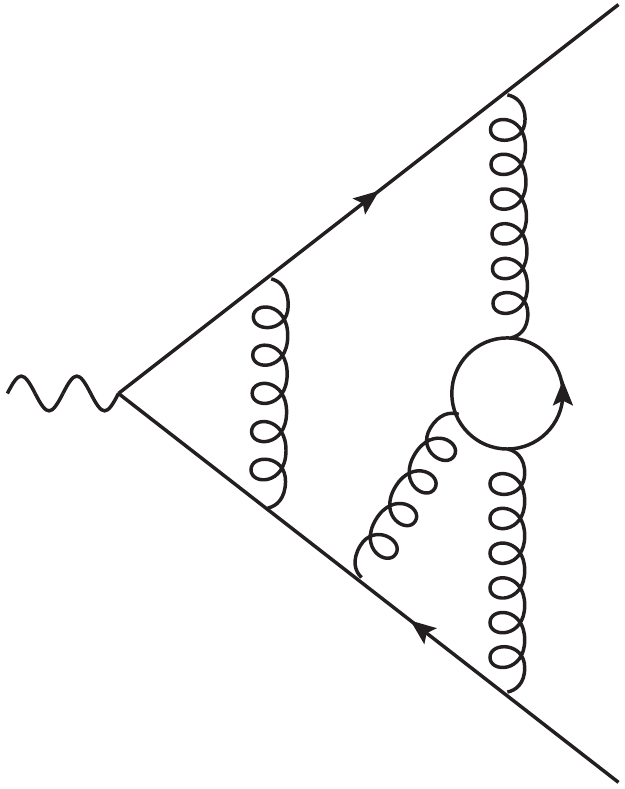}
\FDiag{$n_f C_F C_A^2$}{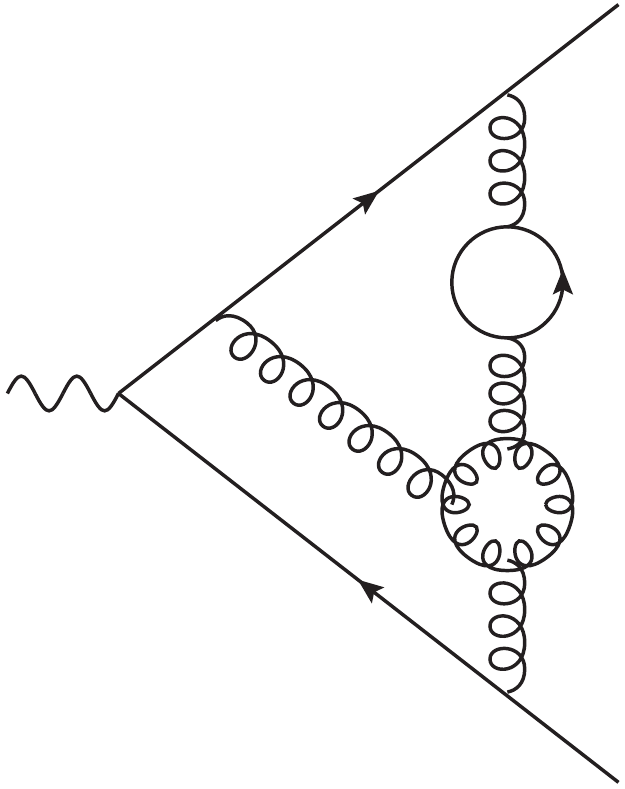}
\FDiag{$n_f (d^{abcd}_F)^2$}{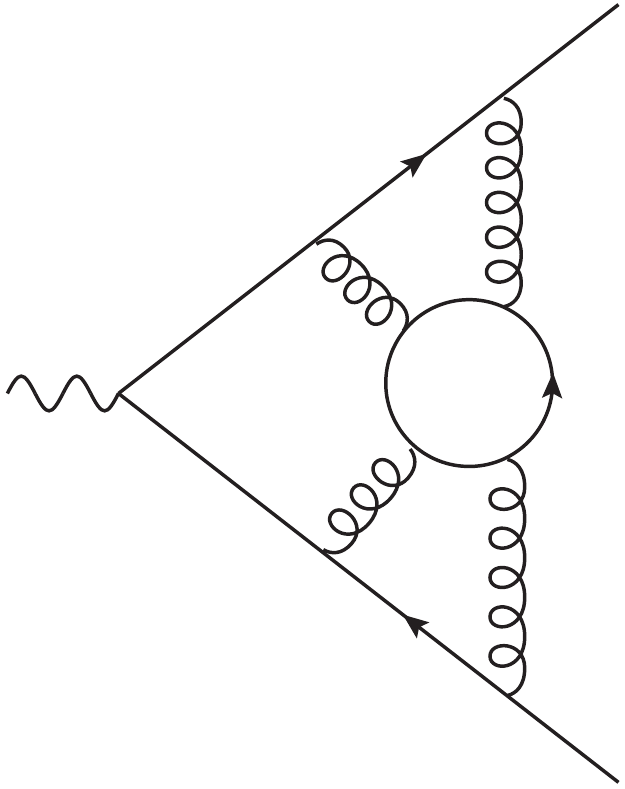}
\\[1ex]
\FDiag{$n_f^2 C_F^2$}{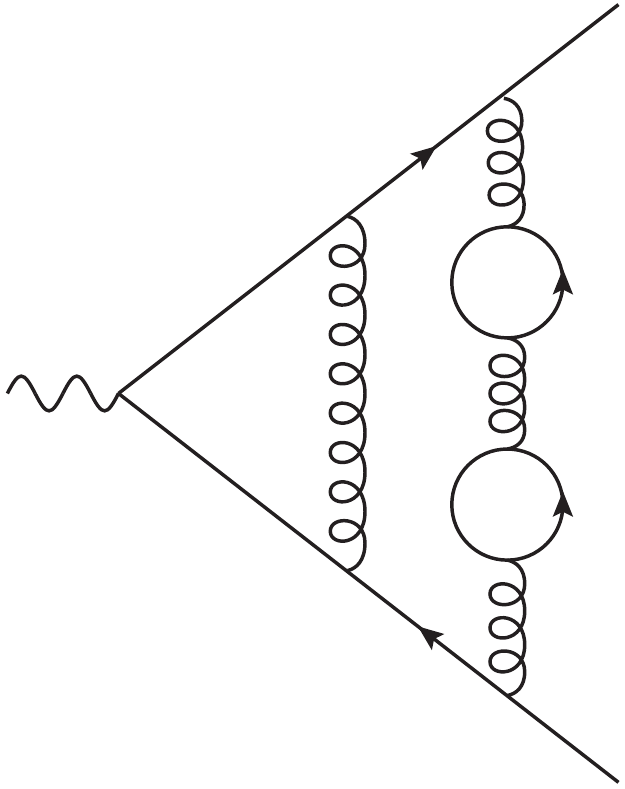}
\FDiag{$n_f^2 C_F C_A$}{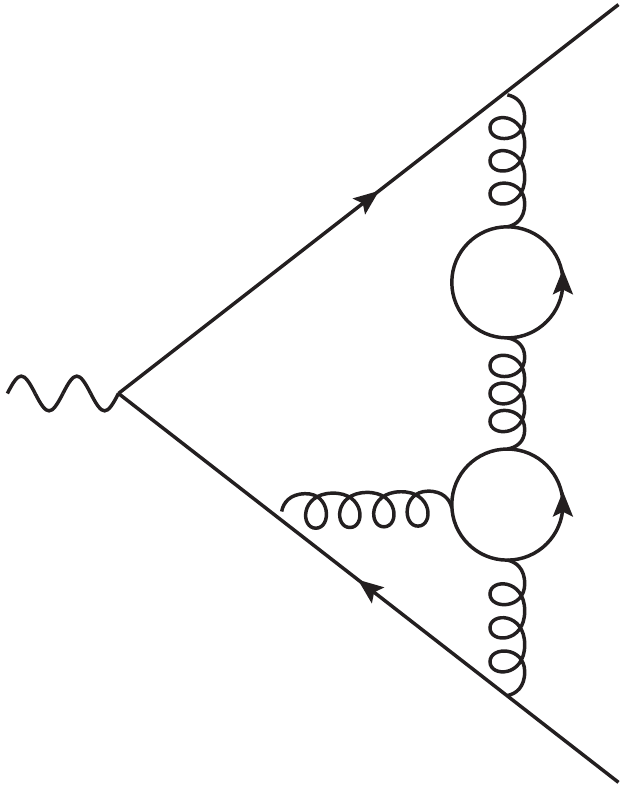}
\FDiag{$n_f^3 C_F$}{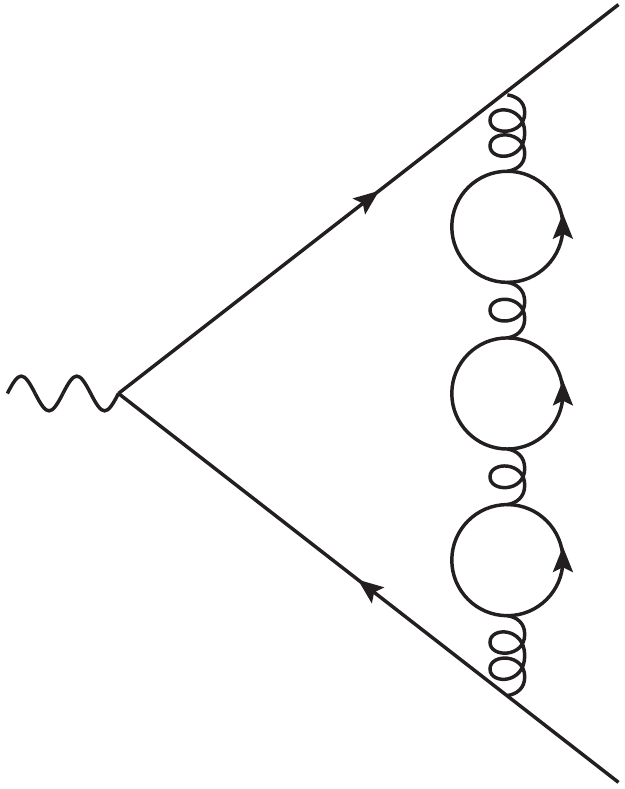}
\FDiag{$n_{q\gamma} (d^{abc}_F)^2 C_F$}{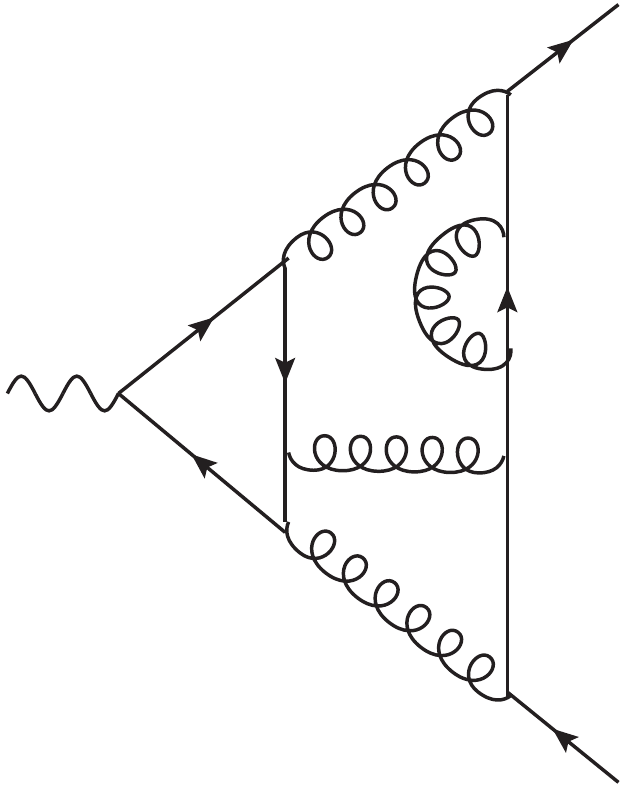}
\FDiag{$n_{q\gamma} (d^{abc}_F)^2 C_A$}{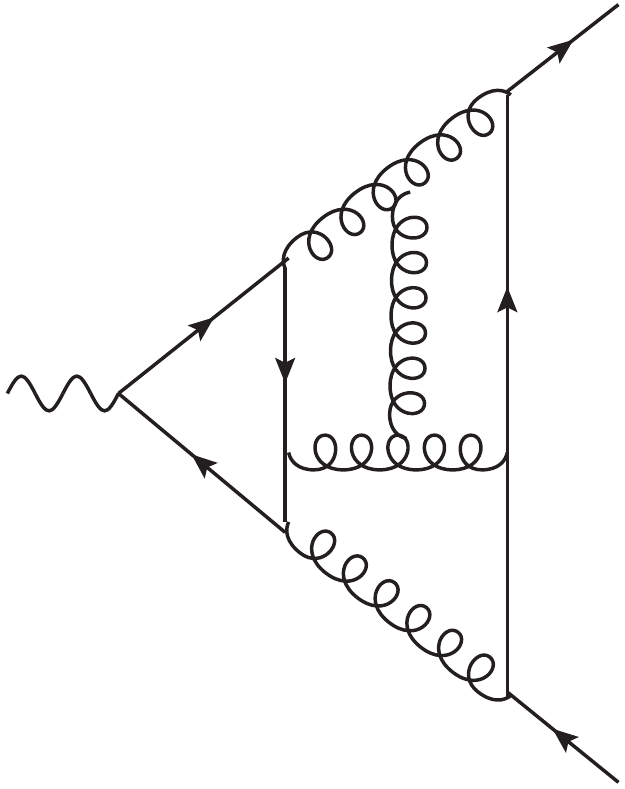}
\FDiag{$n_{q\gamma}n_f  (d^{abc}_F)^2$}{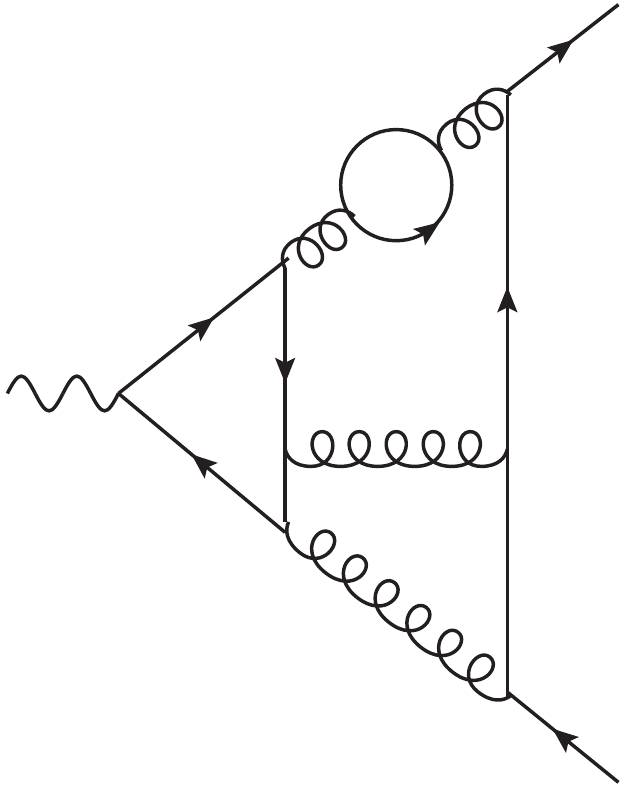}
\\[2ex]
\FDiag{$n_f C_A^3$}{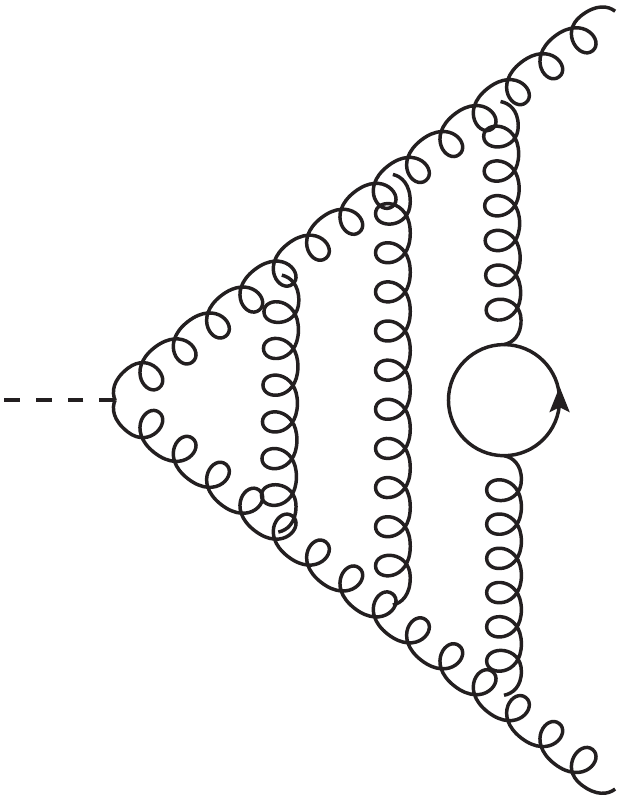}
\FDiag{$n_f C_A^2 C_F$}{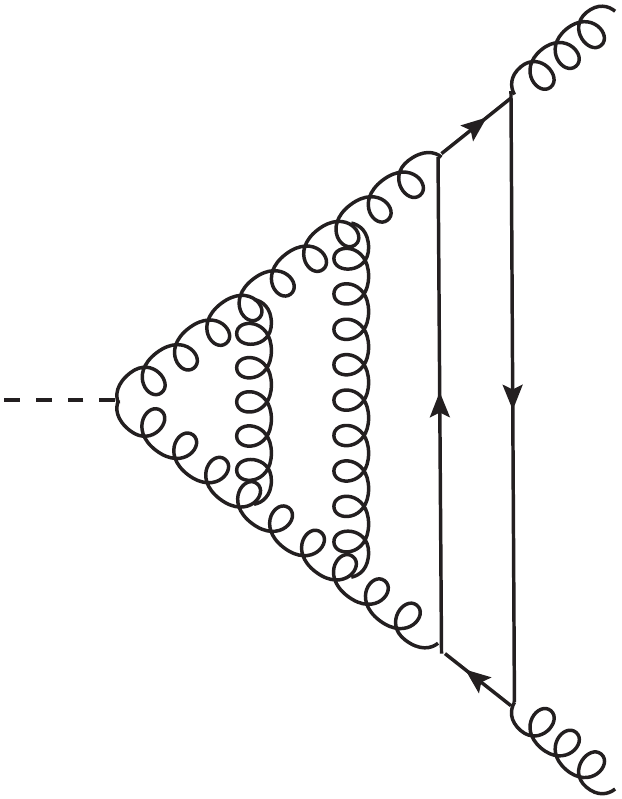}
\FDiag{$n_f C_A C_F^2$}{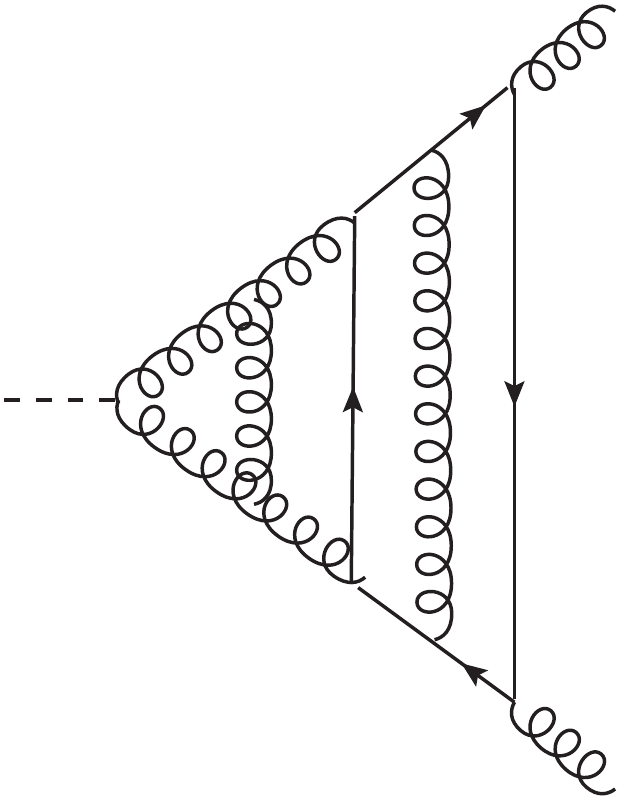}
\FDiag{$n_f C_F^3$}{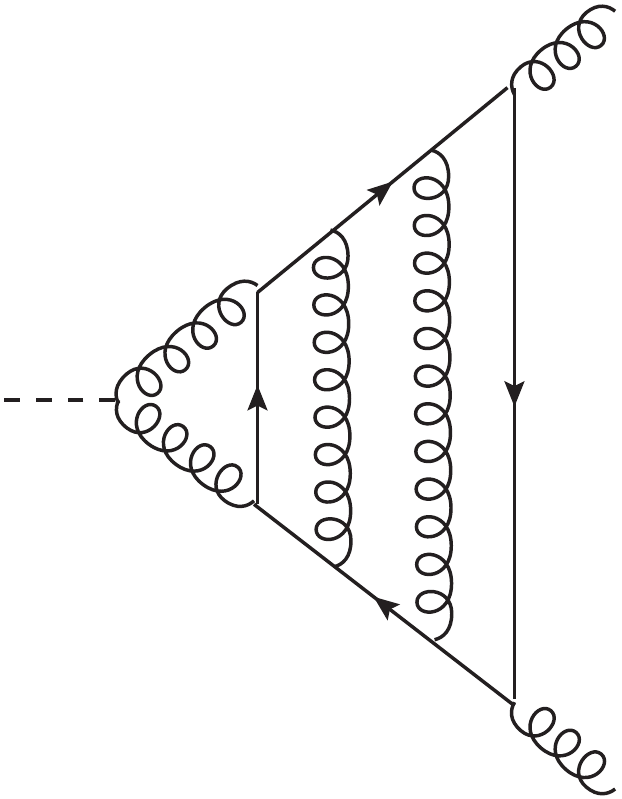}
\FDiag{\hspace*{-2ex}$n_f d_A^{abcd}d_F^{abcd}$}{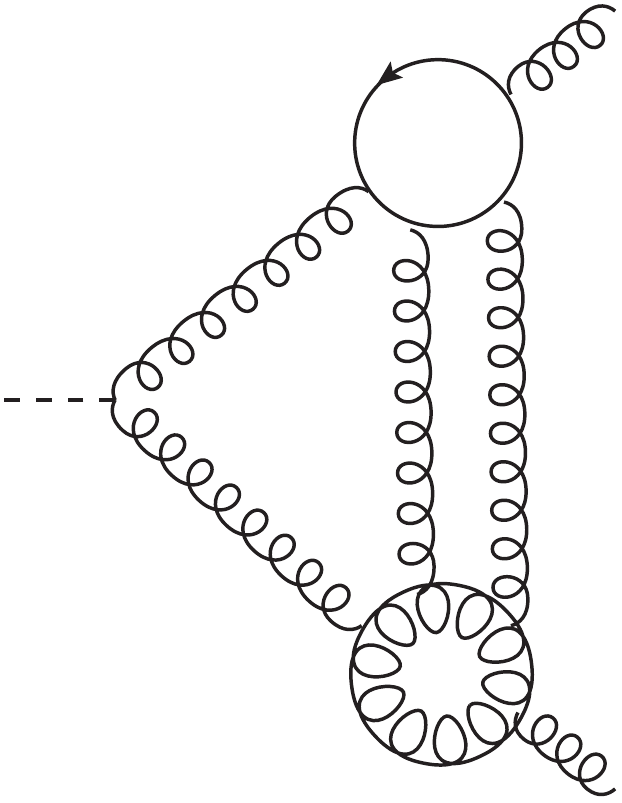}
\\[1ex]
\FDiag{$n_f^2 C_A^2$}{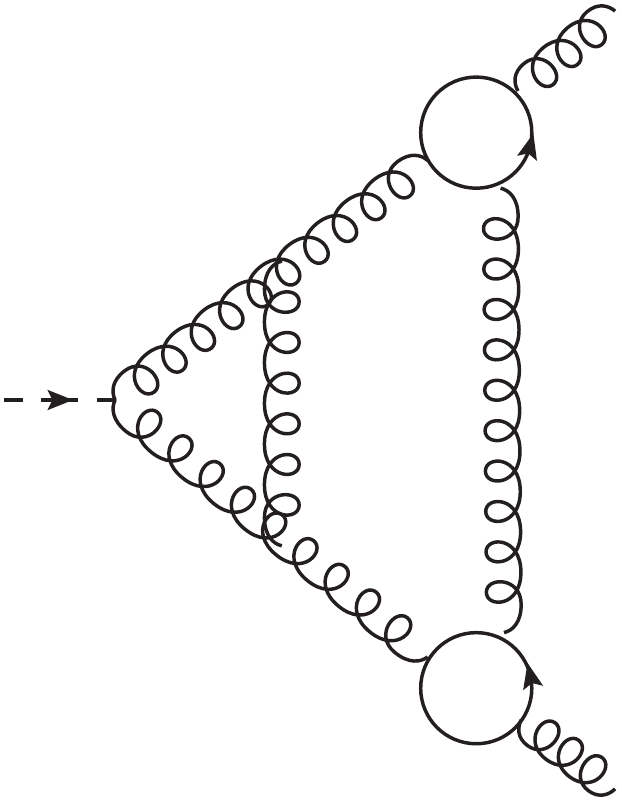}
\FDiag{$n_f^2 C_A C_F$}{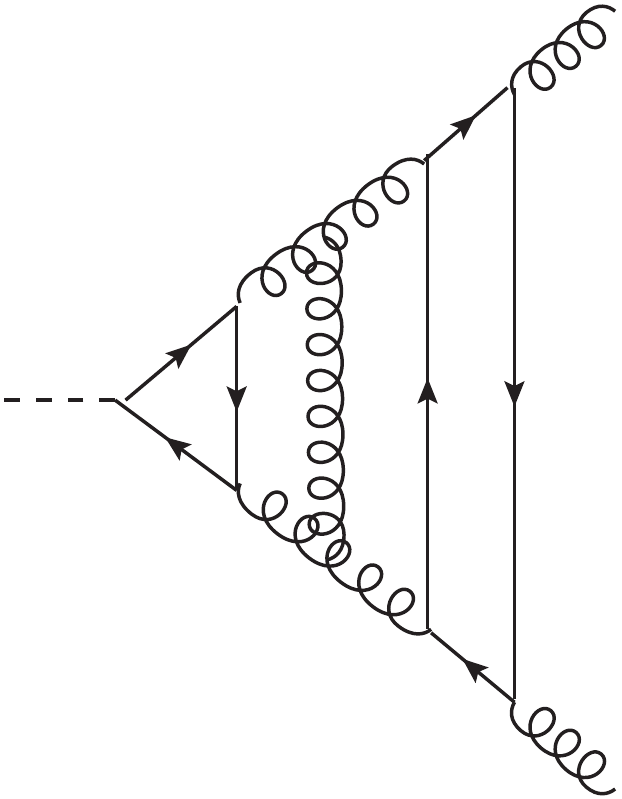}
\FDiag{$n_f^2 C_A C_F^2$}{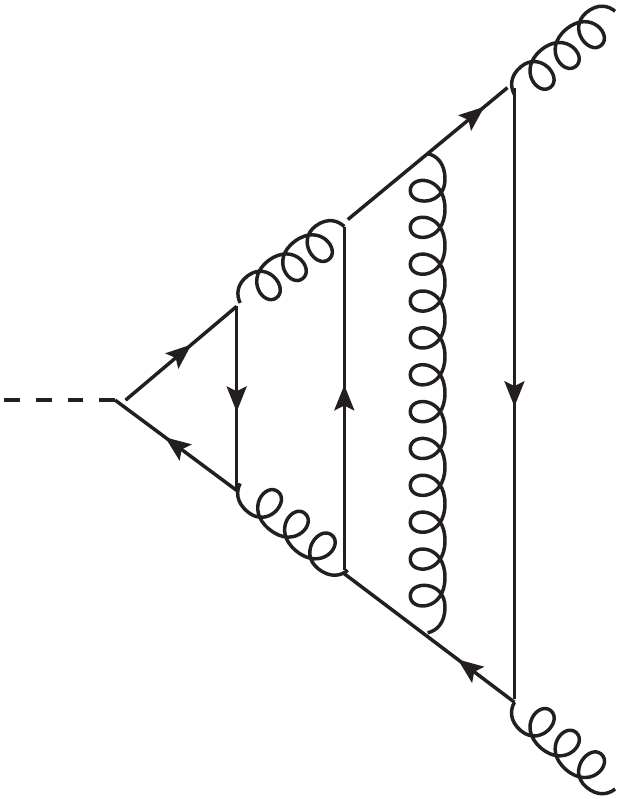}
\FDiag{$n_f^2 (d_F^{abcd})^2$}{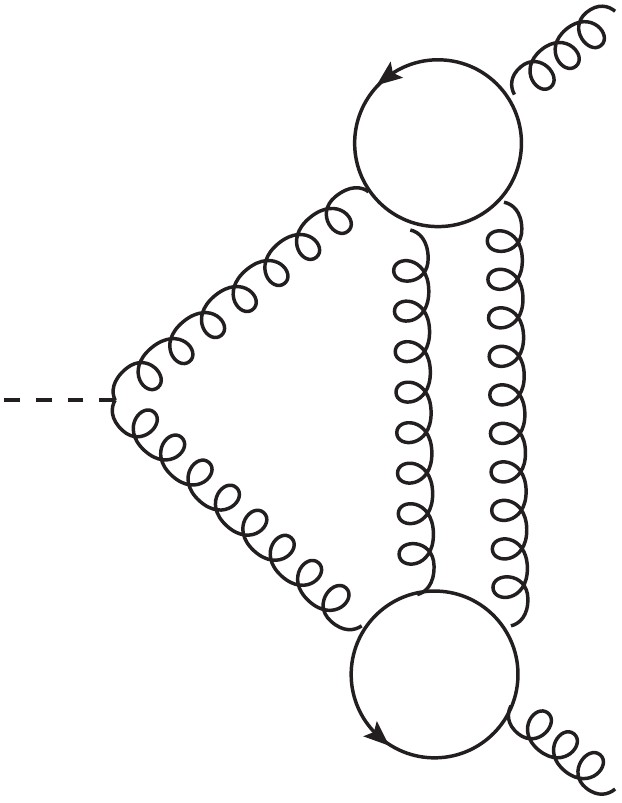}
\FDiag{$n_f^3 C_A$}{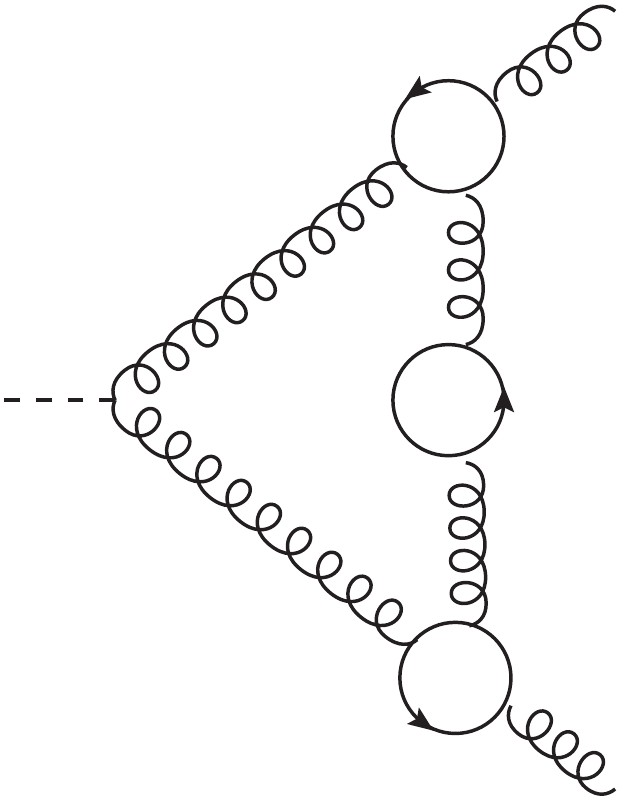}
\FDiag{$n_f^3 C_F$}{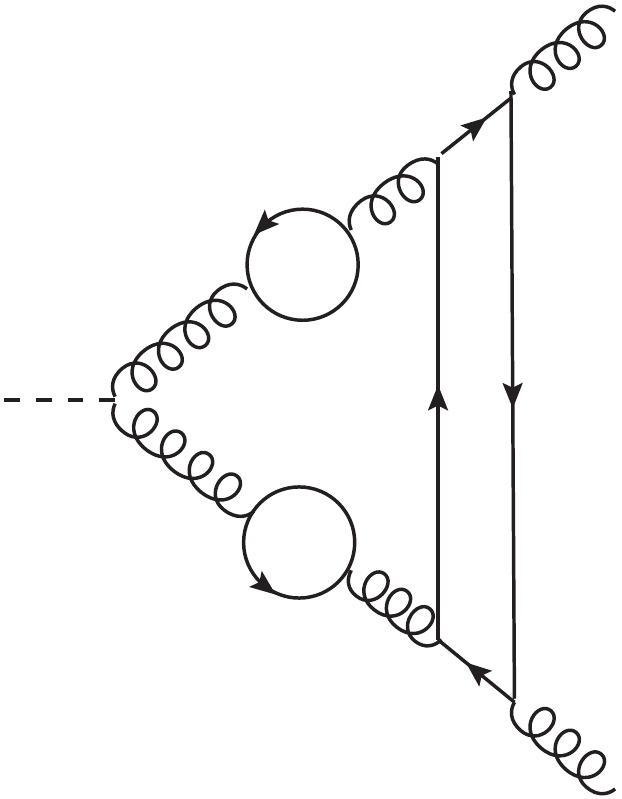}
\caption{\label{fig::diags}Sample Feynman diagrams contributing to the
  fermionic corrections to $F_q$ and $F_g$
  at four-loop order. Straight and curly lines denote
  quarks and gluons, respectively.
  Both planar and non-planar diagrams contribute.
}
\end{center}
\end{figure*}

The virtual corrections are conveniently expressed in terms of form factors of
the photon-quark vertex and the effective gluon-Higgs boson vertex. Let us denote
the corresponding bare vertex functions by $\Gamma^{\mu}_q$ and
$\Gamma^{\mu\nu}_g$, respectively. Then the bare form factors are obtained
from
\begin{eqnarray}
  F_q(q^2) &=& -\frac{1}{4(1-\epsilon)q^2}
  \mbox{Tr}\left( q_2\!\!\!\!\!/\,\,\, \Gamma^\mu q_1\!\!\!\!\!/\,\,\,
    \gamma_\mu\right)
  \,,
  \nonumber\\
  F_g(q^2) &=&
  \frac{\left(q_1\cdot q_2\,\,
      g_{\mu\nu}-q_{1,\mu}\,q_{2,\nu}-q_{1,\nu}\,q_{2,\mu}\right)}
  {2(1-\epsilon)}
  \Gamma^{\mu\nu}_g
  \,,
\end{eqnarray}
where our overall normalization is such that both form factors are one at leading order.
Further, we work in conventional dimensional regularization
and use $d=4-2\epsilon$ for the space-time dimension. The external momentum of the
photon and Higgs is $q=q_1+q_2$ and $q_1$ and $q_2$ are the incoming momenta
of the quark and anti-quark in the case of $F_q$ and of the
gluons in the case of $F_g$.
Some sample Feynman diagrams contributing to the fermionic part of
$F_q$ and $F_g$ are shown in Fig.~\ref{fig::diags}.

We define the perturbative expansion
of $F_q$ and $F_g$ in terms of the bare strong coupling constant
and write
\begin{eqnarray}
  F_x &=& 1 +
          \sum_{n\ge1} 
          \left(\frac{\alpha_s^0}{4\pi}\right)^n
         \left(\frac{4\pi}{e^{\gamma_E}}\right)^{n\epsilon}
          \left(\frac{\mu^2}{-q^2-i0} \right)^{n\epsilon}
          F_x^{(n)}
          \,,\nonumber\\
  \label{eq::FFbare}
\end{eqnarray}
with $x\in\{q,g\}$.

Two-loop corrections to $F_q$ have been computed in
Refs.~\cite{Kramer:1986sg,Matsuura:1987wt,Matsuura:1988sm,Gehrmann:2005pd} and
the first two-loop calculation for $F_g$ has been performed
in~\cite{Harlander:2000mg}. In the first three-loop calculation of $F_q$ and
$F_g$~\cite{Baikov:2009bg} the coefficients of the highest $\epsilon$
expansion terms of three master integrals were only known numerically.
These coefficients have been computed in~\cite{Lee:2010ik}. The results
of~\cite{Baikov:2009bg} have been confirmed in
Refs.~\cite{Gehrmann:2010ue,Gehrmann:2010tu,vonManteuffel:2015gxa}. 
For the computation of three-loop master integrals we also refer to~\cite{Heinrich:2009be}.

At four-loop order there are only partial results.
For $F_q$, the large-$N_c$ limit,
which only involves planar diagrams, has been considered in
Refs.~\cite{Henn:2016men,Lee:2016ixa},
the $n_f^2$ terms are available from~\cite{Lee:2017mip},
the complete contribution from color structure $(d_F^{abcd})^2$ has been computed
in~\cite{Lee:2019zop} and confirmed in \cite{vonManteuffel:2020vjv}.
For $F_q$ and $F_g$, all corrections with three or two closed fermion loops were calculated in  \cite{vonManteuffel:2016xki,vonManteuffel:2019wbj}, respectively, including also the singlet contributions.

There are a number of works where pole parts of the form factors have been
computed. In fact, from the $1/\epsilon^2$ pole it is possible to extract the
so-called cusp anomalous dimension. A complete calculation based on the form
factors can be found in Ref.~\cite{vonManteuffel:2020vjv}. In that work a basis
of finite integrals has been chosen and expanded only to lower orders in $\ep$
in order to obtain the required weight six information.
Ref.~\cite{vonManteuffel:2020vjv} confirmed the expression presented in Ref.~\cite{Henn:2019swt},
which is based on a calculation in $\mathcal{N}=4$ super Yang-Mills, other known QCD results,
and conjectural input for one term in the matter contributions.
Partial contributions to the QCD cusp anomalous dimension are available
from~\cite{Gracey:1994nn,Beneke:1995pq,vonManteuffel:2016xki,Henn:2016men,Lee:2016ixa,Lee:2017mip,Davies:2016jie,Grozin:2018vdn}
and numerical results are presented in Refs.~\cite{Moch:2017uml,Moch:2018wjh}.

Recently, the collinear anomalous dimensions of the quark and gluon form
factors have been computed in Ref.~\cite{Agarwal:2021zft} by extracting the
$1/\epsilon$ poles of the corresponding form factors. All contributions
could be computed analytically except one contribution from a 
non-planar four-loop integral defined in $6-2\epsilon$ dimensions,
which is parameterized by ${\cal H}$ (c.f.\ Eq.~(10) of~\cite{Agarwal:2021zft}),
\begin{equation}
\oset[-.6ex]{{(6-2\epsilon)\qquad}}{\Graph{.15}{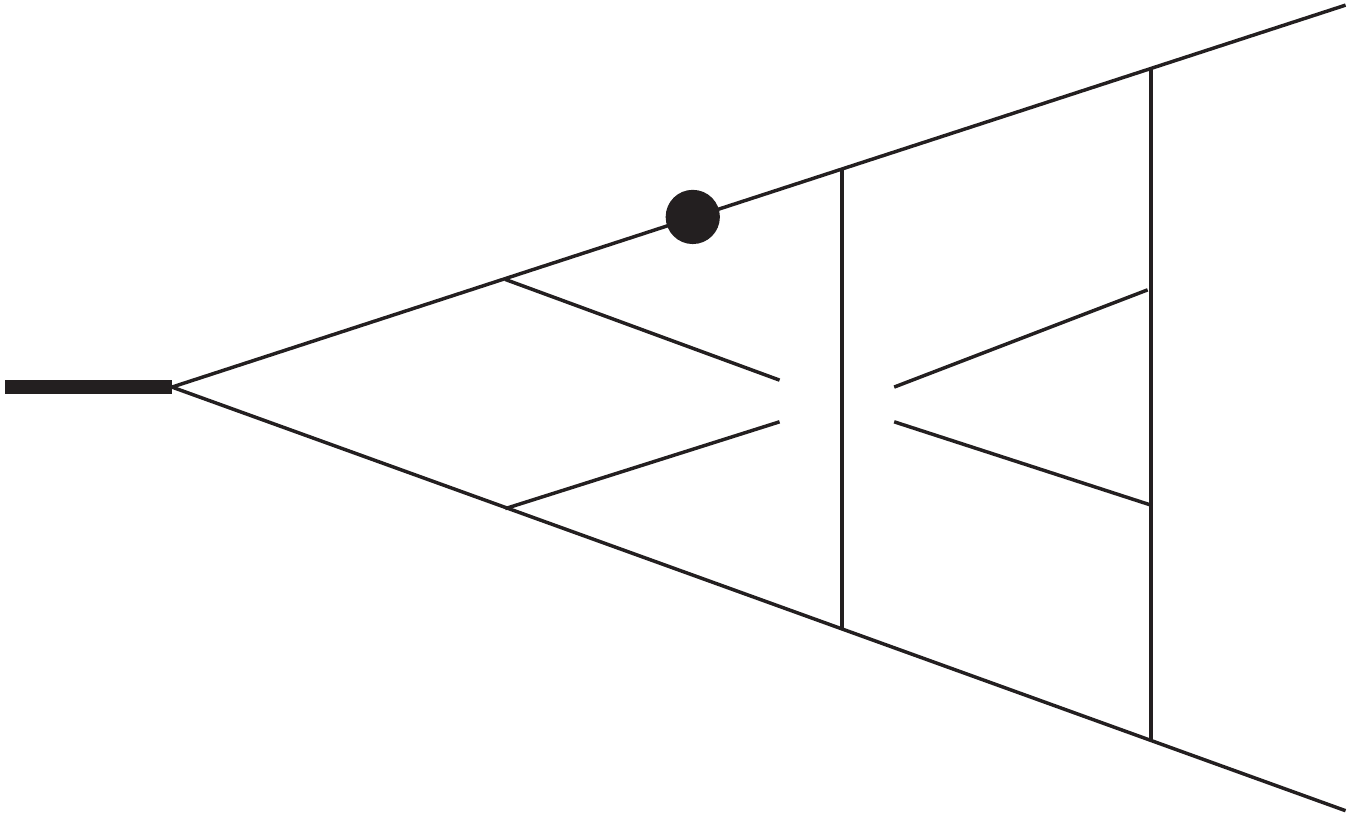}} \equiv \mathcal{H} + {\mathcal{O}\left(\epsilon\right)}.
\end{equation}
Using a numerical evaluation of ${\cal H}$ to 10 significant digits
together with an assumption on the multiple zeta values present in ${\cal H}$,
an analytic expression could be conjectured. Using the results obtained in the
present paper we confirm this expression; details are presented in the next 
Section where we outline some of the techniques used for our calculation.
Our analytic results are presented in Section~\ref{sec::res} and
Section~\ref{sec::con} contains our conclusions and an outlook to the
full result.


\section{\label{sec::calc}Calculation}

We employ \texttt{Qgraf}~\cite{Nogueira:1991ex} to generate the required
Feynman diagrams with closed fermion loops, 2464 diagrams for $F_q^{(4)}$
and 18642 diagrams for $F_g^{(4)}$.
After applying the projectors and performing the numerator algebra with \texttt{Form\;4}~\cite{Kuipers:2012rf}, we obtain the form factors
$F_q$ and $F_g$ expressed as a linear combination of scalar functions belonging to properly defined integral families.
Each function has 18 indices where up to twelve correspond to the
different propagators of the diagrams.
In addition to planar diagrams (see, {\it e.g.}, Ref.~\cite{Henn:2016men,Lee:2016ixa,vonManteuffel:2019gpr}) there are
non-planar diagrams; it is the latter which pose challenges.
We perform the calculation in a general $R_\xi$ gauge
and check explicitly that terms proportional to $\xi$ cancel in our results.

From the computational point of view there are two challenges one has to deal
with. The first one is the integration-by-parts reduction~\cite{Tkachov:1981wb,Chetyrkin:1981qh} of the scalar integrals, which appear in
the amplitudes, to so-called master integrals.
For this task, we employ the setup described in Ref.~\cite{vonManteuffel:2020vjv},
which is based on the codes \texttt{Reduze\;2}~\cite{vonManteuffel:2012np} and \texttt{Finred}, implementing techniques from~\cite{Laporta:2001dd,vonManteuffel:2014ixa,Gluza:2010ws,Boehm:2017wjc,Lee:2013mka,Bitoun:2017nre,Agarwal:2020dye}.

The second challenge is the computation of the master integrals as a Laurent
series in $\epsilon$.  Here we have
two approaches at hand. The first one is based on the construction of a basis
of finite master integrals~\cite{vonManteuffel:2014qoa,vonManteuffel:2015gxa,Schabinger:2018dyi}, partly in $6-2\epsilon$ dimensions. Subsequently
the program {\tt HyperInt}~\cite{Panzer:2014caa} is used to compute the
$\epsilon$ expansion of the master integrals.
This approach allowed us to compute all $\epsilon$ coefficients of the master
integrals required for the fermionic four-loop corrections except for $\mathcal{H}$.
We wish to note, however, that it remains unclear whether the evaluation of the constants of
transcendental weight eight (or even higher) of some of the non-planar twelve-line
master integrals is possible in this approach.
In particular, for the two Feynman integrals corresponding to non-planar
graphs with twelve edges
\begin{equation}
\label{eq:nonlinred}
   \Graph{.15}{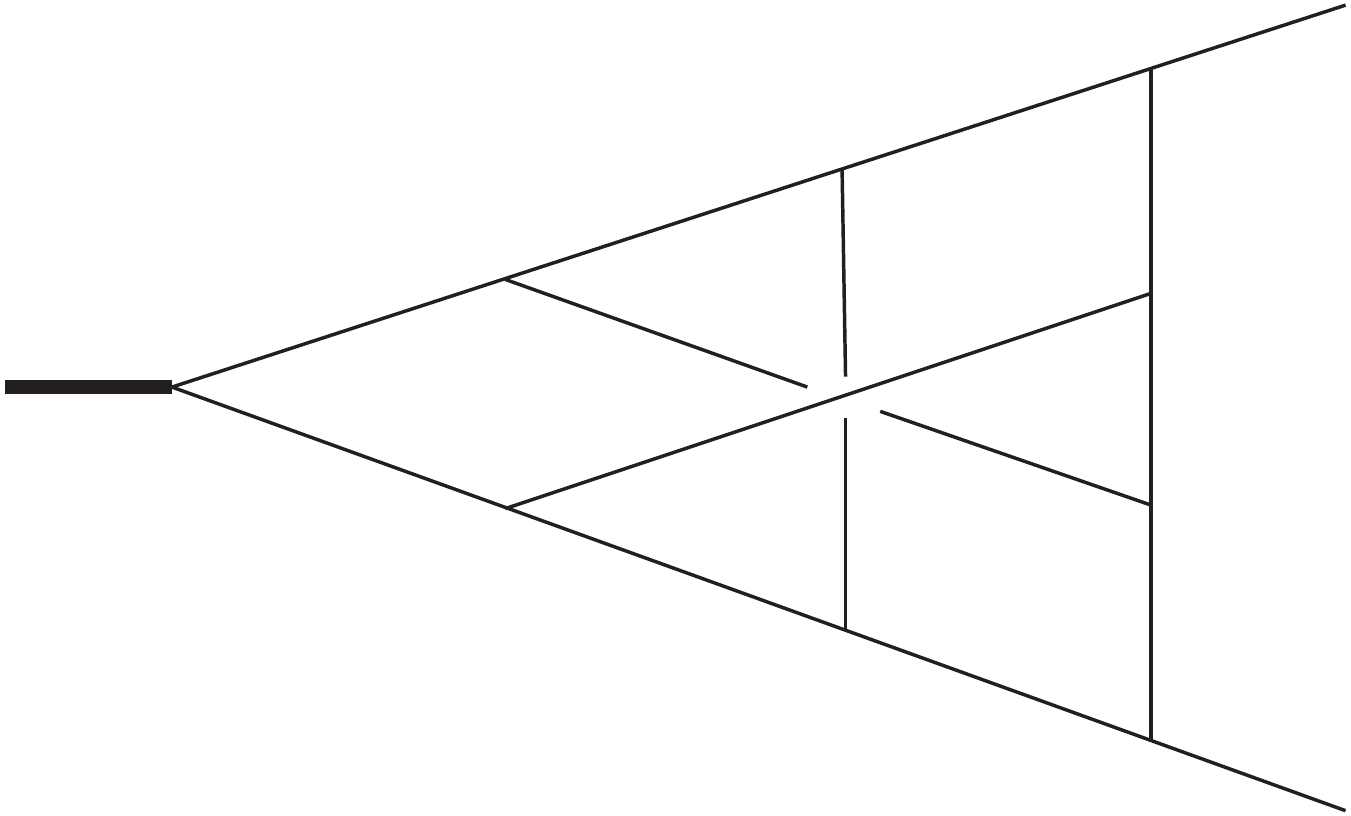}  \qquad \Graph{.15}{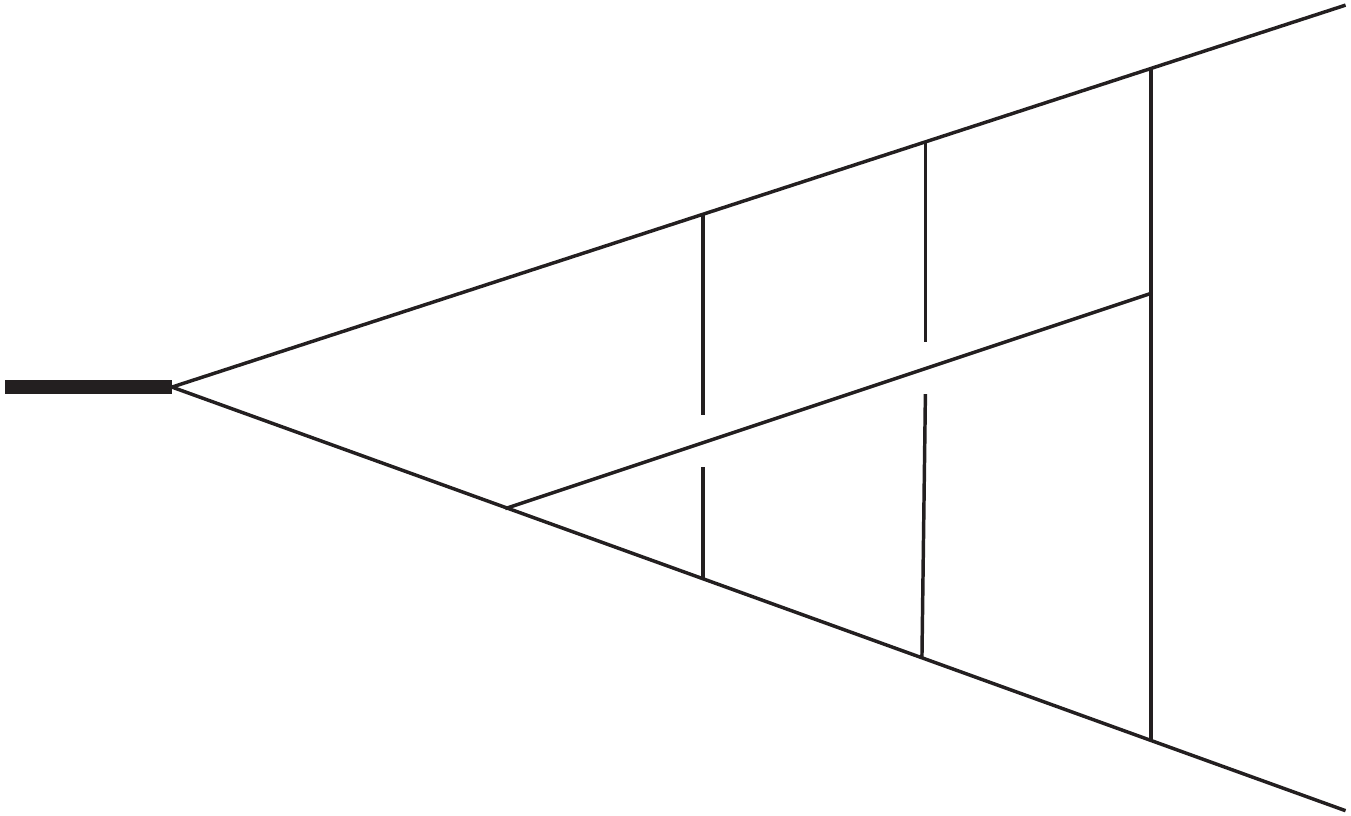}
\end{equation}
it is not known whether a linearly reducible~\cite{Brown:2008um,Brown:2009ta} Feynman
parametric representation exists.

In this Letter, we show that both remaining non-linearly reducible topologies
\eqref{eq:nonlinred} can be solved using a second method, which is based on differential equations.
While the method of differential equations~\cite{Kotikov:1990kg,Bern:1993kr,Gehrmann:1999as} is not directly applicable to one-scale Feynman integrals, we can introduce an
additional scale parameter, as was suggested in Ref.~\cite{Henn:2013nsa}. On the one hand, we complicate the situation. On the other hand, we obtain the possibility to apply the full power of the method of differential equations (see, {\it e.g.}, Refs.~\cite{Henn:2013pwa,Henn:2014qga} and  Ref. \cite[Section E.8]{Blondel:2018mad}).
In the context of massless four-loop form factors this approach has been applied successfully in
Refs.~\cite{Henn:2016men,Lee:2016ixa,Lee:2017mip,Lee:2019zop}.  In a first step
one introduces a second mass scale by imposing a virtuality $q_2^2\not=0$ on
one of the external partons, which apparently makes the problem more
complicated. However, we are now in a position to establish
differential equations for the master integrals in the variable $x =q_2^2/q^2$
which determine the connection between the points $x=1$ and $x=0$. The boundary conditions are then easy to fix at $x=1$ as in this point our integrals turn into massless propagator integrals for which analytic results are known at least up to weight twelve~\cite{Baikov:2010hf,Lee:2011jt}.
A detailed description of the procedure can be found in Ref.~\cite{Lee:2019zop}.

In our calculation we employ \texttt{Fire\;6}~\cite{Smirnov:2019qkx} in combination with \texttt{LiteRed}~\cite{Lee:2013mka,Lee:2012cn} to compute the reductions for
the differential equations and closely follow the algorithm of
Refs.~\cite{Lee:2014ioa,Lee2017c} as implemented in \texttt{Libra}~\cite{,Lee:2020zfb} to bring our
system in $\epsilon$-form. The complexity of the two topologies \eqref{eq:nonlinred} is somewhat reflected also in the properties of the differential equations. First, it appears that differential equations for these two topologies, in addition to singularities at $x=0$ and $1$, have singularities at  $x=-1,\, 4,\,1/4$ (at $x=4$) for the first (second) topology, respectively. Among those, the singularity at $x=1/4$ is especially troublesome as it lies inside the segment $(0,1)$ connecting the points of interest. Moreover, it appears that in order to reduce the system to $\epsilon$-form, we need to introduce the algebraic extensions $x_1=\sqrt{x}$, $x_2=\sqrt{x-1/4}$, and $x_3=\sqrt{1/x-1/4}$. Fortunately, for each specific iterated integral which appears in the $\epsilon$ expansion of the master integrals of these two topologies it is possible to find the rationalizing variable change. As a result, the master integrals of the two topologies are not directly expressed via non-alternating multiple zeta values, but rather via Goncharov polylogarithms with letters in the alphabet $\{0,\pm1,\pm i\sqrt{3}, e^{\pm i \pi/3}, e^{\pm 2 i \pi/3}, e^{\pm i \pi/3}/2 \}$. Using the PSLQ algorithm, we were able to express our final result for the master integrals with two massless legs and their subtopologies through to weight nine in terms of regular zeta values, $\zeta_n$ $(n=2,\ldots,9)$, and
the multiple zeta value
\begin{equation}
    \zeta_{5,3} = \sum_{m=1}^\infty \sum_{n=1}^{m-1} \frac{1}{m^5 n^3} \approx 0.0377076729848\,.
\end{equation} 
Our results for the corner integrals of the two non-planar topologies through to the finite parts are
\begin{widetext}
\begin{align}
\label{eq:D27631corner}
&\Graph{.15}{D_12_27631}=
\frac{1}{\ep^8} \Big( \mfrac{7}{18} \Big)
+ \frac{1}{\ep^7} \Big( \mfrac{55}{24} \Big)
+ \frac{1}{\ep^6} \Big(
    - \mfrac{67}{9} \zeta_2
    - \mfrac{797}{144}
    \Big)
+ \frac{1}{\ep^5}\Big(
    - \mfrac{442}{9} \zeta_3
    - \mfrac{643}{18} \zeta_2
    + \mfrac{1193}{144}
    \Big)
+ \frac{1}{\ep^4}\Big(
    - \mfrac{9199}{360} \zeta_2^2
    - \mfrac{3547}{18} \zeta_3
\nonumber \\ & \qquad
    + \mfrac{7793}{72} \zeta_2
    + \mfrac{1013}{48}
    \Big)
+ \frac{1}{\ep^3}\Big(
    - \mfrac{2858}{3} \zeta_5
    + \mfrac{27617}{36} \zeta_3 \zeta_2 
    - \mfrac{3439}{180} \zeta_2^2
    + \mfrac{60893}{72} \zeta_3
    - \mfrac{1897}{8} \zeta_2
    - \mfrac{43895}{144}
    \Big)
+ \frac{1}{\ep^2}\Big(
    \mfrac{179927}{72} \zeta_3^2
    - \mfrac{40853}{252} \zeta_2^3
\nonumber \\ & \qquad
    - \mint{2780} \zeta_5
    + \mfrac{23467}{9} \zeta_3 \zeta_2 
    + \mfrac{132359}{180} \zeta_2^2
    - \mfrac{66607}{24} \zeta_3
    - \mfrac{5423}{72} \zeta_2
    + \mfrac{311383}{144}
    \Big)
+ \frac{1}{\ep}\Big(
    - \mfrac{1015395}{32} \zeta_7
    + \mfrac{30493}{2} \zeta_5 \zeta_2
    + \mfrac{274199}{90} \zeta_3 \zeta_2^2
\nonumber \\ & \qquad
    + \mfrac{44984}{9} \zeta_3^2
    - \mfrac{540823}{420} \zeta_2^3
    + \mfrac{477281}{24} \zeta_5
    - \mfrac{412181}{36} \zeta_3 \zeta_2 
    - \mfrac{117101}{30} \zeta_2^2
    + \mfrac{410629}{72} \zeta_3
    + \mfrac{400999}{72} \zeta_2
    - \mfrac{622069}{48}
    \Big)
+ \mfrac{122261}{15} \zeta_{5,3}
\nonumber \\ & \qquad
+ \mfrac{1298525}{12} \zeta_5 \zeta_3
- \mfrac{942899}{36} \zeta_3^2 \zeta_2 
- \mfrac{121150681}{9000} \zeta_2^4
- \mfrac{2558101}{16} \zeta_7
+ \mfrac{360793}{6} \zeta_5 \zeta_2
- \mfrac{53821}{18} \zeta_3 \zeta_2^2
- \mfrac{1428953}{72} \zeta_3^2
+ \mfrac{2037031}{168} \zeta_2^3
\nonumber \\ & \qquad
- \mfrac{1989461}{24} \zeta_5
+ \mfrac{526387}{12} \zeta_3 \zeta_2 
+ \mfrac{245017}{18} \zeta_2^2
+ \mfrac{738547}{72} \zeta_3
- \mfrac{1198061}{24} \zeta_2
+ \mfrac{10519199}{144}
+ \mathcal{O}(\ep),
\\[1ex]
\label{eq:E47063corner}
&\Graph{.15}{E_12_47063} =
\frac{1}{\ep^8} \Big( \mfrac{1}{18} \Big)
+ \frac{1}{\ep^7} \Big( \mfrac{13}{48} \Big)
+ \frac{1}{\ep^6} \Big(
  - \mfrac{4}{9}\zeta_2
  - \mfrac{193}{192}
 \Big)
+ \frac{1}{\ep^5} \Big(
  - \mfrac{395}{72}\zeta_3
  - \mfrac{175}{144}\zeta_2
  + \mfrac{167}{48}
 \Big)
+ \frac{1}{\ep^4} \Big(
  - \mfrac{13577}{720}\zeta_2^2
  - \mfrac{347}{18}\zeta_3
\nonumber \\ & \qquad
  + \mfrac{373}{96}\zeta_2
  - \mfrac{385}{36}
 \Big)
+ \frac{1}{\ep^3} \Big(
  - \mfrac{5941}{24}\zeta_5
  - \mfrac{799}{36}\zeta_3 \zeta_2
  - \mfrac{130661}{1440}\zeta_2^2
  + \mfrac{1043}{32}\zeta_3
  - \mfrac{1225}{72}\zeta_2
  + \mfrac{1771}{72}
 \Big)
+ \frac{1}{\ep^2} \Big(
  - \mfrac{1687}{72}\zeta_3^2
  - \mfrac{1495399}{5040}\zeta_2^3
\nonumber \\ & \qquad
  - \mfrac{23419}{24}\zeta_5
  - \mfrac{6953}{24}\zeta_3 \zeta_2
  + \mfrac{13921}{90}\zeta_2^2
  - \mfrac{6655}{72}\zeta_3
  + \mfrac{730}{9}\zeta_2
  + \mfrac{11}{4}
 \Big)
+ \frac{1}{\ep^1} \Big(
  - \mfrac{433869}{64}\zeta_7
  - \mfrac{21667}{24}\zeta_5 \zeta_2
  + \mfrac{101723}{90}\zeta_3 \zeta_2^2
  - \mfrac{5267}{12}\zeta_3^2
\nonumber \\ & \qquad
  - \mfrac{14629019}{10080}\zeta_2^3
  + \mfrac{257675}{96}\zeta_5
  + \mfrac{30845}{144}\zeta_3 \zeta_2
  - \mfrac{981}{8}\zeta_2^2
  + \mfrac{45509}{72}\zeta_3
  - \mfrac{8633}{24}\zeta_2
  - \mfrac{41461}{72}
 \Big)
  + \mfrac{43916}{15}\zeta_{5,3}
  + \mfrac{44957}{4}\zeta_5 \zeta_3
\nonumber \\ & \qquad
  + \mfrac{133033}{36}\zeta_3^2 \zeta_2
  - \mfrac{257784979}{56000}\zeta_2^4
  - \mfrac{1611457}{64}\zeta_7
  - \mfrac{207481}{24}\zeta_5 \zeta_2
  + \mfrac{4033391}{720}\zeta_3 \zeta_2^2
  + \mfrac{846605}{288}\zeta_3^2
  + \mfrac{4194707}{1440}\zeta_2^3
  - \mfrac{76189}{12}\zeta_5
\nonumber \\ & \qquad
  + \mfrac{61787}{36}\zeta_3 \zeta_2
  - \mfrac{168493}{360}\zeta_2^2
  - \mfrac{59869}{12}\zeta_3
  + \mfrac{48311}{36}\zeta_2
  + \mfrac{197863}{36}
+\mathcal{O}(\ep)
\end{align}
\end{widetext} 
in the conventions of~\cite{vonManteuffel:2019gpr}.
Combining the integral solutions obtained by direct integration with the result \eqref{eq:D27631corner}
allows us to determine
\begin{align}
\mathcal{H} &=
    \mfrac{161}{16} \zeta_7
    + \mfrac{5}{2} \zeta_5 \zeta_2
    - \frac{5}{2} \zeta_3 \zeta_2^2
    + \mint{10} \zeta_3^2
    + \mfrac{223}{210} \zeta_2^3
    - \mint{25} \zeta_5
\nonumber \\ &\quad
    - \mint{6} \zeta_3 \zeta_2
    + \mfrac{3}{10} \zeta_2^2
    + \mint{9} \zeta_3\,.
\end{align} 
This value for $\mathcal{H}$ agrees with the expression conjectured
in \cite{Agarwal:2021zft} and thus confirms the non-fermionic contributions to
the collinear anomalous dimensions in that reference analytically.
Moreover, this result provides the last remaining master integral coefficient
required for the present calculation.


\section{\label{sec::res}Results for form factors}

In this Section we present the complete fermionic four-loop corrections to
the form factors $F_q$ and $F_g$ in massless QCD. We express the results in 
terms of $SU(N_c)$ color factors and use
\begin{gather}
  C_F=\frac{N_c^2-1}{2N_c}\,,\quad C_A=N_c\,, \quad N_A=N_c^2-1\,,
  \nonumber\\
  N_F=N_c\,,\quad
  n_{q\gamma} = \sum_{q^\prime} \frac{Q_{q^\prime}}{Q_q}\,,\quad
  \frac{d_F^{abc}d_F^{abc}}{N_A} = \frac{N_c^2 - 4}{16 N_c} \,,
  \nonumber\\
  \frac{d_F^{abcd}d_F^{abcd}}{N_A} = \frac{N_c^4 - 6 N_c^2 + 18}{96 N_c^2} \,,
  \nonumber\\
  \frac{d_A^{abcd}d_F^{abcd}}{N_A} = \frac{N_c(N_c^2 + 6)}{48} \,,
      \label{eq::d44}
\end{gather}
where $Q_q$ is the fractional charge of the quark $q$ and
$n_f$ is the number of active quark flavors. Without loss of generality we
have used for the trace normalization $T_F=1/2$.

The $\epsilon$ expansion of the fermionic corrections to both form factors start
with seventh order poles in $1/\epsilon$, reflecting the fact that fermionic corrections
start to contribute only at two loops.
Similarly, four-loop contributions with more than two closed fermion loops or specific color factors
start even later in the $\ep$ expansion.
The corresponding poles through to order $1/\ep$ can be obtained from \cite{vonManteuffel:2020vjv}; they consist of zeta values with transcendental weight up to six.

Here, we calculate the complete finite part of the fermionic four-loop contributions to $F_q$ and $F_g$ and obtain an analytical result in terms of zeta values with transcendental weight up to seven.
%
Our result for the finite part of $F_q^{(4)}$ reads
\begin{widetext}
\begin{align}
\left. F_q^{(4)} \right|_{\epsilon^0} &=
\CS{n_f C_F^3} \Big(
    \mfrac{1153615}{126} \zeta_7
    + \mfrac{6316}{9} \zeta_5 \zeta_2
    + \mfrac{229468}{135} \zeta_3 \zeta_2^2
    + \mfrac{547270}{81} \zeta_3^2
    + \mfrac{1341628}{2835} \zeta_2^3
    - \mfrac{3467995}{162} \zeta_5
    - \mfrac{192737}{81} \zeta_3 \zeta_2
    - \mfrac{1420133}{1215} \zeta_2^2
\notag\\ &
    - \mfrac{38482147}{972} \zeta_3
    + \mfrac{12734681}{648} \zeta_2
    + \mfrac{7837713013}{419904}
    \Big)
+ \CS{n_f C_A C_F^2 } \Big(
    \mfrac{5669}{4} \zeta_7
    - \mfrac{249194}{135} \zeta_5 \zeta_2
    - \mfrac{417244}{405} \zeta_3 \zeta_2^2
    - \mfrac{11438080}{729} \zeta_3^2
\notag\\ &
    - \mfrac{38510}{63} \zeta_2^3
    + \mfrac{4959127}{243} \zeta_5
    + \mfrac{408107}{81} \zeta_3 \zeta_2
    + \mfrac{8388679}{2916} \zeta_2^2
    + \mfrac{2642551543}{26244} \zeta_3
    - \mfrac{111491363}{2187} \zeta_2
    - \mfrac{326984889779}{3779136}
    \Big)
\notag\\ &
+ \CS{n_f C_A^2 C_F } \Big(
    \mfrac{6943}{24} \zeta_7
    + \mfrac{2755}{9} \zeta_5 \zeta_2
    + \mfrac{1912}{45} \zeta_3 \zeta_2^2 
    + \mfrac{202210}{27} \zeta_3^2
    + \mfrac{128953}{1620} \zeta_2^3
    - \mfrac{34844257}{3240} \zeta_5
    - \mfrac{119489}{108} \zeta_3 \zeta_2
    - \mfrac{1251893}{540} \zeta_2^2
\notag\\ &
    - \mfrac{111467677}{1944} \zeta_3
    + \mfrac{123861583}{3888} \zeta_2
    + \mfrac{21075909203}{279936}
    \Big)
+ \CS{n_f \mfrac{d_F^{abcd}d_F^{abcd} }{N_F}} \Big(
    - \mint{1240} \zeta_7
    + \mfrac{992}{3} \zeta_5 \zeta_2
    - \mfrac{3952}{15} \zeta_3 \zeta_2^2  
    + \mfrac{680}{9} \zeta_3^2
    + \mfrac{41620}{189} \zeta_2^3
\notag\\ &
    + \mfrac{95098}{27} \zeta_5
    + \mfrac{92}{3} \zeta_3 \zeta_2 
    + \mfrac{7552}{45} \zeta_2^2
    - \mfrac{21566}{9} \zeta_2
    - \mfrac{13414}{27} \zeta_3
    + \mfrac{3190}{3}
    \Big)
+ \CS{n_f^2 C_F^2 } \Big(
    \mfrac{689582}{729} \zeta_3^2
    + \mfrac{191252}{945} \zeta_2^3
    + \mfrac{187364}{1215} \zeta_5
\notag\\ &
    + \mfrac{177800}{729} \zeta_3 \zeta_2
    + \mfrac{4777}{135} \zeta_2^2
    - \mfrac{90719803}{13122} \zeta_3
    + \mfrac{44208841}{13122} \zeta_2
    + \mfrac{5325319081}{944784}
    \Big)
+ \CS{n_f^2 C_A C_F } \Big(
    - \mfrac{1714}{3} \zeta_3^2
    - \mfrac{2836}{315} \zeta_2^3
    + \mfrac{150886}{135} \zeta_5
\notag\\ &
    - \mfrac{436}{3} \zeta_3 \zeta_2
    - \mfrac{3722}{135} \zeta_2^2
    + \mfrac{2897315}{486} \zeta_3
    - \mfrac{5825827}{1296} \zeta_2
    - \mfrac{3325501813}{279936}
    \Big)
+ \CS{n_f^3 C_F } \Big(
    - \mfrac{2194}{135} \zeta_5
    - \mfrac{820}{81} \zeta_3 \zeta_2
    + \mfrac{3322}{135} \zeta_2^2
    - \mfrac{20828}{243} \zeta_3
\notag\\ &
    + \mfrac{145115}{729} \zeta_2
    + \mfrac{10739263}{17496}
    \Big)
+ \CS{n_{q\gamma}\mfrac{d_F^{abc}d_F^{abc}C_F }{N_F}} \Big(
    \mint{26624} \zeta_7
    + \mint{1792} \zeta_5 \zeta_2
    + \mfrac{35584}{15} \zeta_3 \zeta_2^2
    + \mfrac{30688}{9} \zeta_3^2
    - \mfrac{179152}{189} \zeta_2^3
    - \mfrac{170224}{27} \zeta_5
\notag\\ &
    - \mfrac{8656}{3} \zeta_3 \zeta_2
    + \mfrac{60416}{45} \zeta_2^2
    - \mfrac{100624}{9} \zeta_2
    - \mfrac{71552}{27} \zeta_3
    - \mfrac{89360}{9}
    \Big)
+ \CS{n_{q\gamma} \mfrac{d_F^{abc}d_F^{abc}C_A}{N_F} } \Big(
    - \mfrac{13972}{3} \zeta_7
    - \mint{1840} \zeta_5 \zeta_2 
    - \mfrac{784}{5} \zeta_3 \zeta_2^2
\notag\\ &
    - \mint{13008} \zeta_3^2
    - \mfrac{618328}{189} \zeta_2^3
    + \mfrac{54436}{9} \zeta_5
    + \mfrac{15112}{3} \zeta_3 \zeta_2
    - \mfrac{95692}{45} \zeta_2^2
    + \mfrac{45976}{9} \zeta_3
    + \mfrac{107984}{9} \zeta_2
    + \mfrac{65264}{9}
    \Big)
+  \CS{n_{q\gamma} n_f \mfrac{d_F^{abc}d_F^{abc} }{N_F}} \Big(
    \mint{2304} \zeta_3^2
\notag\\ &
    + \mfrac{545792}{945} \zeta_2^3
    - \mfrac{8512}{9} \zeta_5
    - \mfrac{1888}{3} \zeta_3 \zeta_2
    + \mfrac{12448}{45} \zeta_2^2
    - \mfrac{3520}{9} \zeta_3
    - \mfrac{16928}{9} \zeta_2
    - \mfrac{11296}{9}
    \Big)
\notag\\ &
+ \text{contributions without closed fermion loop}.
\end{align}
%
For the finite part of $F_g^{(4)}$ we obtain
\begin{align}
\left. F_g^{(4)}\right|_{\ep^0} &=
\CS{n_f C_A^3} \Big(
    \mfrac{365579}{63} \zeta_7
    - \mfrac{64151}{135} \zeta_5 \zeta_2
    + \mfrac{410228}{405} \zeta_3 \zeta_2^2
    - \mfrac{11278261}{2916} \zeta_3^2
    - \mfrac{1672838}{2835} \zeta_2^3
    - \mfrac{24219919}{4860} \zeta_5
    + \mfrac{638345}{486} \zeta_3 \zeta_2
\notag\\ &
    - \mfrac{1464209}{29160} \zeta_2^2
    + \mfrac{483257171}{52488} \zeta_3
    - \mfrac{55297501}{34992} \zeta_2
    - \mfrac{322439904151}{7558272}
    \Big)
+ \CS{n_f C_A^2 C_F} \Big(
    - \mfrac{66967}{36} \zeta_7
    + \mfrac{10022}{9} \zeta_5 \zeta_2
    - \mfrac{8824}{9} \zeta_3 \zeta_2^2
\notag\\ &
    + \mfrac{129887}{81} \zeta_3^2
    - \mfrac{163538}{945} \zeta_2^3
    + \mfrac{2166853}{1620} \zeta_5
    + \mfrac{91172}{81} \zeta_3 \zeta_2
    - \mfrac{10414}{27} \zeta_2^2
    + \mfrac{25799269}{5832} \zeta_3
    - \mfrac{6154445}{3888} \zeta_2
    - \mfrac{821501405}{139968}
    \Big)
\notag\\ &
+ \CS{n_f C_A C_F^2} \Big(
    \mint{2692} \zeta_7
    - \mint{1420} \zeta_5 \zeta_2
    + \mfrac{2500}{3} \zeta_3 \zeta_2^2
    + \mfrac{2720}{3} \zeta_3^2
    + \mfrac{376624}{945} \zeta_2^3
    - \mfrac{50591}{6} \zeta_5
    + \mfrac{1321}{3} \zeta_3 \zeta_2
    - \mfrac{4057}{5} \zeta_2^2
    + \mfrac{85357}{36} \zeta_3
\notag\\ &
    + \mfrac{6451}{36} \zeta_2
    - \mfrac{24275}{432}
    \Big)
+ \CS{n_f C_F^3} \Big(
    \mint{3360} \zeta_7
    - \mint{2940} \zeta_5
    - \mint{156} \zeta_3
    + \mfrac{169}{2}
    \Big)
+ \CS{n_f \mfrac{d_A^{abcd}d_F^{abcd}}{N_A}} \Big(
    \mfrac{2464}{3} \zeta_7
    + \mint{1824} \zeta_5 \zeta_2
    - \mfrac{1088}{3} \zeta_3 \zeta_2^2
\notag\\ &
    - \mfrac{15700}{3} \zeta_3^2
    - \mfrac{245536}{945} \zeta_2^3
    + \mfrac{108692}{9} \zeta_5
    + \mfrac{1544}{9} \zeta_3 \zeta_2
    - \mfrac{35108}{45} \zeta_2^2
    - \mfrac{89932}{9} \zeta_3
    + \mfrac{9580}{27} \zeta_2
    + \mfrac{6944}{9}
    \Big)
+ \CS{n_f^2 C_A^2} \Big(
    \mfrac{717266}{729} \zeta_3^2
\notag\\ &
    + \mfrac{58858}{945} \zeta_2^3
    - \mfrac{62671}{1215} \zeta_5
    - \mfrac{87329}{1458} \zeta_3 \zeta_2
    + \mfrac{380701}{3240} \zeta_2^2
    + \mfrac{27036815}{52488} \zeta_3
    - \mfrac{53253361}{104976} \zeta_2
    + \mfrac{110016540845}{7558272}
    \Big)
+ \CS{n_f^2 C_A C_F} \Big(
    \! - \mfrac{4354}{3} \zeta_3^2
\notag\\ &
    - \mfrac{143524}{945} \zeta_2^3
    - \mfrac{47948}{27} \zeta_5
    - \mfrac{3688}{9} \zeta_3 \zeta_2
    - \mfrac{13820}{81} \zeta_2^2
    - \mfrac{538853}{162} \zeta_3
    - \mfrac{27629}{81} \zeta_2
    + \mfrac{3697777}{324}
    \Big)
+ \CS{n_f^2 C_F^2 } \Big(
    \mfrac{2432}{3} \zeta_3^2
    + \mfrac{32512}{135} \zeta_2^3
\notag\\ &
    + \mint{3360} \zeta_5
    + \mfrac{128}{3} \zeta_3 \zeta_2
    - \mfrac{484}{3} \zeta_2^2
    - \mfrac{36260}{9} \zeta_3
    + \mfrac{439}{9} \zeta_2
    - \mfrac{80281}{216}
    \Big)
+ \CS{n_f^2 \mfrac{d_F^{abcd}d_F^{abcd}}{N_A}} \Big(
    \mint{512} \zeta_3^2
    - \mint{960} \zeta_5
    + \mfrac{384}{5} \zeta_2^2
    + \mint{1520} \zeta_3
\notag\\ &
    - \mfrac{9008}{9}
    \Big)
+ \CS{n_f^3 C_A} \Big(
    - \mfrac{14474}{135} \zeta_5
    + \mfrac{4556}{81} \zeta_3 \zeta_2
    - \mfrac{1418}{27} \zeta_2^2
    - \mfrac{99890}{243} \zeta_3
    + \mfrac{38489}{729} \zeta_2
    - \mfrac{20832641}{17496}
    \Big)
+ \CS{n_f^3 C_F} \Big(
    - \mfrac{508069}{324}
\notag\\ &
    + \mfrac{4288}{27} \zeta_5
    - \mint{64} \zeta_3 \zeta_2
    + \mfrac{2288}{27} \zeta_2^2
    + \mfrac{24812}{27} \zeta_3
    + \mfrac{3074}{27} \zeta_2
    \Big)
\notag\\ &
+ \text{contributions without closed fermion loop}.
\end{align}

\end{widetext}

The $n_f^3$, $n_f^2$ and $n_{q\gamma}n_f$ terms agree
with the results presented in Refs.~\cite{vonManteuffel:2016xki,vonManteuffel:2019wbj}.
Our expression for $F_q$ reproduces the result of Ref.~\cite{Henn:2016men}
in the large-$N_c$ limit.
The remaining, subleading color terms for $F_q^{(4)}$ and all of the terms linear in $n_f$ for $F_g^{(4)}$ are new.


\section{\label{sec::con}Conclusions}

In this Letter, we calculated the complete fermionic corrections
to the photon-quark and Higgs-gluon vertices in massless four-loop QCD.
We solved two non-planar vertex topologies using the method
of differential equations and found a result in terms of multiple zeta
values.
This renders the only two topologies which were not known to be
linearly reducible accessible, such that
the main obstacle for the remaining four-loop corrections has been removed.
Our calculation confirms a previous conjecture for the analytical solution
of one of the integrals in this topology, which fully establishes the
pole terms of all non-fermionic four-loop corrections analytically.


\section*{\label{sec::ack}Acknowledgments}

AvM and RMS gratefully acknowledge Erik Panzer for related collaborations.
This research was supported by the Deutsche Forschungsgemeinschaft (DFG,
German Research Foundation) under grant 396021762 — TRR 257 ``Particle Physics
Phenomenology after the Higgs Discovery'' and by the National Science Foundation (NSF) under grant 2013859 ``Multi-loop amplitudes and precise predictions for the LHC''.
The work of AVS and VAS was supported by the Russian Science Foundation, agreement no.\ 21-71-30003.
We acknowledge the High Performance Computing Center at Michigan State University
for computing resources.
The Feynman diagrams were drawn with the help of {\tt Axodraw}~\cite{Vermaseren:1994je}
and {\tt JaxoDraw}~\cite{Binosi:2003yf}.


%
%


\bibliography{ffml_nf}

\end{document}